\documentclass[]{ws-ijmpa}
\usepackage{amsmath}
\usepackage{amssymb}

\begin{document}

\title{Quantum D-branes and exotic smooth $\mathbb{R}^{4}$%
\thanks{This is part 2 of the work based on the talk ,,Small exotic smooth
$\mathbb{R}^{4}$ and string theory'' given at the International Congress
of Mathematicians, ICM2010, 19-28.08.2010, Hyderabad, India%
} }

\author{Torsten Asselmeyer-Maluga}%
\address{German Aerospace center, Rutherfordstr. 2, 12489 Berlin \\ torsten.asselmeyer-maluga@dlr.de%
} 
\author{Jerzy Kr\'ol}%
\address{University of Silesia, Institute of Physics, ul. Uniwesytecka 4, 40-007
Katowice \\ iriking@wp.pl %
}

\maketitle
\begin{history}
\received{Day Month Year}
\revised{Day Month Year}
\end{history}

\begin{abstract}
In this paper, we present the idea that the formalism of string theory
is connected with the dimension 4 in a new way, not covered by phenomenological
or model-building approaches. The main connection is given by structures
induced by small exotic smooth $\mathbb{R}^{4}$'s having intrinsic
meaning for physics in dimension 4. We extend the notion of stable
quantum D-branes in a separable noncommutative $C^{\star}$ algebras
over convolution algebras corresponding to the codimension-1 foliations
of $S^{3}$ which are mainly connected to small exotic $\mathbb{R}^{4}$.
The tools of topological K-homology and K-theory as well KK-theory
describe stable quantum branes in the $C^{\star}$ algebras when naturally
extended to algebras. In case of convolution algebras, small exotic
smooth $\mathbb{R}^{4}$'s embedded in exotic $\mathbb{R}^{4}$ correspond
to a generalized quantum branes on the algebras. These results extend
the correspondence between exotic $\mathbb{R}^{4}$ and classical
D and NS branes from our previous work. 

\keywords{exotic ${\mathbb R}^4$; quantum D-branes; qunatum D-branes in $C^{\star}$-algebras.}
\end{abstract}
%\tableofcontents{}

\section{Introduction}

In this paper we further explore the relation between string theory
and some 4-dimensional structures such that these structures do not
appear as a result of usual compactification or model-buildings in
string theory. These structures are rather involved in the mathematics
of string theory but they are able to encode (in 4 dimensions) some
dynamics of branes configuration and the geometry of certain string
backgrounds. At the same time, these structures are of fundamental
importance for 4-dimensional physics. 

In our previous paper \cite{AsselmeyerKrol2011} we observed a correlation
between D as well NS brane configurations in some backgrounds and
the appearance of exotic smoothness on the topological $\mathbb{R}^{4}$.
It is known that the $\mathbb{R}^{4}$ with standard smoothness structure
is part of the string background. A variation of the brane configurations
induce a change of the smoothness structure, i.e. one has to consider
different smoothings of the $\mathbb{R}^{4}$. But this result is
the unique feature of $\mathbb{R}^{n}$ holding only for $n=4$ where
a variety (actually a continuum) of smoothings of $\mathbb{R}^{n}$
must exist. In fact in any other dimension $n\neq4$ there exists
precisely a unique standard smooth $\mathbb{R}^{n}$ \cite{Asselmeyer2007}.
Physics corresponding to exotic smooth $\mathbb{R}^{4}$ has been
gradually exhibited since the nineties. In a recent series of papers,
new aspects important for quantum gravity are being worked out \cite{AsselmeyerKrol2009,AsselmeyerKrol2009a,AsselmeyerKrol2010,Krol2010}
directly.

The recognition of the role of exotic $\mathbb{R}^{4}$ in string
theory relies so far on the following steps:
\begin{itemize}
\item Standard smooth $\mathbb{R}^{4}$appears as a part of a exact string
background;
\item The process of changing the exotic smoothness on $\mathbb{R}^{4}$
is capable to encode the change in the configuration of specific D
or NS branes \cite{AsselmeyerKrol2011}.
\item All exotic $\mathbb{R}^{4}$'s appearing in this setup are \emph{small
exotic $\mathbb{R}^{4}$'s}, i.e. a small exotic $\mathbb{R}^{4}$
embeds smoothly in the standard smooth $\mathbb{R}^{4}$ as open subsets. 
\end{itemize}
Thus, string configurations can be expressed inherently in terms of
4-dimensional structures, i.e. exotic smooth $\mathbb{R}^{4}$'s are
complex enough to encode some string configurations. Particularly
all these phenomena disappear when one changes the smoothness to the
standard one. 

In this paper we consider the quantum regime of D-branes in string
theory. Especially the correct setup for quantum branes is an open
problem. However a natural proposal is the consideration of (non-commutative)
$C^{\star}$-algebras replacing (classical, submanifold-like) branes
as well manifold spacetime. In the context of $C^{\star}$-algebras
there are many important counterparts of differential-geometric results
including Poincar\'e duality, characteristic classes or the Riemann-Roch
theorem. Especially one obtains a generalized formula for charges
of quantum D-branes \cite{Szabo2008b,Szabo2008a}.

The basic technical ingredient of the analysis of small exotic $\mathbb{R}^{4}$'s
is the relation between exotic (small) $\mathbb{R}^{4}$'s and non-cobordant
codimension-1 foliations of $S^{3}$ as well gropes and wild embeddings
as shown in \cite{AsselmeyerKrol2009}. The foliation of the 3-sphere
is classified by the Godbillon-Vey class as element of the cohomology
group $H^{3}(S^{3},\mathbb{R})$. By using the $S^{1}$-gerbes it
was possible to interpret the integral elements $H^{3}(S^{3},\mathbb{Z})$
as characteristic classes of a $S^{1}$-gerbe over $S^{3}$ \cite{AsselmeyerKrol2009a}.
In the next section we will explain the whole complex of ideas more
carefully. Then we present some facts and definitions of K-homology
and KK-theory used to introduce stable D-branes as K-theory classes
in terms of tachyons condensation. These K-theory classes can be naturally
described by use of K-string theory (e.g. \cite{AsakawaSugimotoTerasima2002}).
Furthermore there is a canonical interpretation for spectral triples
including tachyon fields. This correspondence is further developed
into the realm of noncommutative $C^{\star}$-algebras, following
e.g. \cite{Szabo2008a,Szabo2008b}, in section \ref{sub:Branes-on-separable}.
Now a natural interpretation of quantum stable D-branes is given by
branes on the $C^{\star}$-algebra. In fact a categorical description
is necessary for an understanding of quantum D-branes: objects are
quantum D-branes and the morphisms in the category are KK-theory classes.
Then in section \ref{sec:Exotic--and-branes} we explore the notion
of stable D-branes in the convolution non-commutative algebra of the
foliations representing exotic $\mathbb{R}^{4}$'s. In section \ref{sub:Net-of-exotic}
we establish the (partial) correspondence between stable D-branes
as above and the net of exotic smooth $\mathbb{R}^{4}$'s embedded
in some exotic $\mathbb{R}^{4}$. A discussion of the results closes
the paper.

\section{Exotic $\mathbb{R}^{4}$ and codimension-one foliations of the 3-sphere}

The main line of the topological argumentation can be briefly described
as follows:
\begin{enumerate}
\item In Bizacas exotic $\mathbb{R}^{4}$ one starts with the neighborhood
$N(A)$ of the Akbulut cork $A$ in the K3 surface $M$. The exotic
$\mathbb{R}^{4}$ is the interior of $N(A)$.
\item This neighborhood $N(A)$ decomposes into $A$ and a Casson handle
representing the non-trivial involution of the cork.
\item From the Casson handle we construct a grope containing Alexanders
horned sphere.
\item Akbuluts construction gives a non-trivial involution, i.e. the double
of that construction is the identity map.
\item From the grope we get a polygon in the hyperbolic space $\mathbb{H}^{2}$.
\item This polygon defines a codimension-1 foliation of the 3-sphere inside
of the exotic $\mathbb{R}^{4}$ with an wildly embedded 2-sphere,
Alexanders horned sphere \cite{Alex:24}.
\item Finally we get a relation between codimension-1 foliations of the
3-sphere and exotic $\mathbb{R}^{4}$.
\end{enumerate}
Now we will explain the details in this construction (see also \cite{AsselmeyerKrol2009}).

An exotic $\mathbb{R}^{4}$ is a topological space with $\mathbb{R}^{4}-$topology
but with a different (i.e. non-diffeomorphic) smoothness structure
than the standard $\mathbb{R}_{std}^{4}$ getting its differential
structure from the product $\mathbb{R}\times\mathbb{R}\times\mathbb{R}\times\mathbb{R}$.
The exotic $\mathbb{R}^{4}$ is the only Euclidean space $\mathbb{R}^{n}$
with an exotic smoothness structure. The exotic $\mathbb{R}^{4}$
can be constructed in two ways: by the failure to arbitrarily split
a smooth 4-manifold into pieces (large exotic $\mathbb{R}^{4}$) and
by the failure of the so-called smooth h-cobordism theorem (small
exotic $\mathbb{R}^{4}$). Here we will use the second method. 

Consider the following situation: one has two topologically equivalent
(i.e. homeomorphic), simple-connected, smooth 4-manifolds $M,M'$,
which are not diffeomorphic. There are two ways to compare them. First
one calculates differential-topological invariants like Donaldson
polynomials \cite{DonKro:90} or Seiberg-Witten invariants \cite{Akb:96}.
But there is another possibility: It is known that one can change
a manifold $M$ to $M'$ by using a series of operations called surgeries.
This procedure can be visualized by a 5-manifold $W$, the cobordism.
The cobordism $W$ is a 5-manifold having the boundary $\partial W=M\sqcup M'$.
If the embedding of both manifolds $M,M'$ in to $W$ induces homotopy-equivalences
then $W$ is called an h-cobordism. Furthermore we assume that both
manifolds $M,M'$ are compact, closed (no boundary) and simply-connected.
As Freedman \cite{Fre:82} showed a h cobordism implies a homeomorphism,
i.e. h-cobordant and homeomorphic are equivalent relations in that
case. Furthermore, for that case the mathematicians \cite{CuFrHsSt:97}
are able to prove a structure theorem for such h-cobordisms:\\
 \emph{Let $W$ be a h-cobordism between $M,M'$. Then there are
contractable submanifolds $A\subset M,A'\subset M'$ together with
a sub-cobordism $V\subset W$ with $\partial V=A\sqcup A'$, so that
the h-cobordism $W\setminus V$ induces a diffeomorphism between $M\setminus A$
and $M'\setminus A'$.} \\
 Thus, the smoothness of $M$ is completely determined (see also
\cite{Akbulut08,Akbulut09}) by the contractible submanifold $A$
and its embedding $A\hookrightarrow M$ determined by a map $\tau:\partial A\to\partial A$
with $\tau\circ\tau=id_{\partial A}$ and $\tau\not=\pm id_{\partial A}$($\tau$
is an involution). One calls $A$, the \emph{Akbulut cork}. According
to Freedman \cite{Fre:82}, the boundary of every contractible 4-manifold
is a homology 3-sphere. This theorem was used to construct an exotic
$\mathbb{R}^{4}$. Then one considers a tubular neighborhood of the
sub-cobordism $V$ between $A$ and $A'$. The interior $int(V)$
(as open manifold) of $V$ is homeomorphic to $\mathbb{R}^{4}$. If
(and only if) $M$ and $M'$ are homeomorphic, but non-diffeomorphic
4-manifolds then $int(V)\cap M$ is an exotic $\mathbb{R}^{4}$. As
shown by Bizaca and Gompf \cite{Biz:94a,BizGom:96} one can use $int(V)$
to construct an explicit handle decomposition of the exotic $\mathbb{R}^{4}$.
We refer for the details of the construction to the papers or to the
book \cite{GomSti:1999}. The idea is simply to use the cork $A$
and add some Casson handle $CH$ to it. The interior of this construction
is an exotic $\mathbb{R}^{4}$. Therefore we have to consider the
Casson handle and its construction in more detail. Briefly, a Casson
handle $CH$ is the result of attempts to embed a disk $D^{2}$ into
a 4-manifold. In most cases this attempt fails and Casson \cite{Cas:73}
looked for a substitute, which is now called a Casson handle. Freedman
\cite{Fre:82} showed that every Casson handle $CH$ is homeomorphic
to the open 2-handle $D^{2}\times\mathbb{R}^{2}$ but in nearly all
cases it is not diffeomorphic to the standard handle \cite{Gom:84,Gom:89}.
The Casson handle is built by iteration, starting from an immersed
disk in some 4-manifold $M$, i.e. a map $D^{2}\to M$ with injective
differential. Every immersion $D^{2}\to M$ is an embedding except
on a countable set of points, the double points. One can kill one
double point by immersing another disk into that point. These disks
form the first stage of the Casson handle. By iteration one can produce
the other stages. Finally consider not the immersed disk but rather
a tubular neighborhood $D^{2}\times D^{2}$ of the immersed disk,
called a kinky handle, including each stage. The union of all neighborhoods
of all stages is the Casson handle $CH$. So, there are two input
data involved with the construction of a $CH$: the number of double
points in each stage and their orientation $\pm$. Thus we can visualize
the Casson handle $CH$ by a tree: the root is the immersion $D^{2}\to M$
with $k$ double points, the first stage forms the next level of the
tree with $k$ vertices connected with the root by edges etc. The
edges are evaluated using the orientation $\pm$. Every Casson handle
can be represented by such an infinite tree. 

The main idea is the construction of a grope, an infinite union of
surfaces with non-vanishing genus, from the Casson handle. But the
grope can be represented by a sequence of polygons in the two-dimensional
hyperbolic space $\mathbb{H}^{2}$. This sequence of polygons is replaced
by one polygon with the same area. From this polygon we can construct
a codimension-one foliation on the 3-sphere as done by Thurston \cite{Thu:72}.
This 3-sphere is part of the boundary $\partial A$ of the Akbulut
cork $A$. Furthermore one can show that the codimension-one foliation
of the 3-sphere induces a codimension-one foliation of $\partial A$
so that the area of the corresponding polygons agree. 

Thus we are able to obtain a relation between an exotic $\mathbb{R}^{4}$
(of Bizaca as constructed from the failure of the smooth h-cobordism
theorem) and codimension-one foliation of the $S^{3}$. Two non-diffeomorphic
exotic $\mathbb{R}^{4}$implying non-cobordant codimension-one foliations
of the 3-sphere described by the Godbillon-Vey class in $H^{3}(S^{3},\mathbb{R})$
(proportional to the are of the polygon). This relation is very strict,
i.e. if we change the Casson handle then we must change the polygon.
But that changes the foliation and vice verse. Finally we obtained
the result:\\
\emph{The exotic $\mathbb{R}^{4}$ (of Bizaca) is determined by
the codimension-1 foliations with non-vanishing Godbillon-Vey class
in $H^{3}(S^{3},\mathbb{R}^{3})$ of a 3-sphere seen as submanifold
$S^{3}\subset\mathbb{R}^{4}$. We say: the exoticness is localized
at a 3-sphere inside the small exotic $\mathbb{R}^{4}$.}

\section{Towards quantum D-branes via K-theory}

In this and subsequent sections we want to show that D-branes of string
theory are related to exotic smooth $\mathbb{R}^{4}$'s also beyond
the semi-classical limit, i.e. in the quantum regime of the theory
where one should deal rather with \emph{quantum branes}. What are
\emph{quantum branes,} is still in general an open and hard problem.
One appealing proposition, relevant for this paper, is to consider
branes in noncommutative spacetimes rather than on commutative manifolds
or orbifolds. This leads to abstract D-branes in general noncommutative
separable $C^{\star}$ algebras as counterparts for quantum D-branes.
The way from D-branes as submanifolds or K-homology classes and spaces
to K-theory cycles, spectral triples and $C^{\star}$ algebras is
presented in the following subsections.

\subsection{D-branes on spaces: K-homology and KK-theory \label{sub:D-branes-on-spaces:} }

The description of systems of stable Dp-branes of IIA,B string theories
via K-theory of topological spaces can be extended toward the branes
in noncommutative $C^{\star}$ algebras. A direct string representation
of the algebraic and K-theoretic ideas can be best explained in K-matrix
string theory where tachyons are elements of the spectral triple representing
the noncommutative geometry of the world-volumes for the configurations
of branes \cite{AsakawaSugimotoTerasima2002}. 

First let us consider the case of a vanishing $H$-field. The charges
of D-branes are classified by suitable K-theory groups, i.e. $K^{0}(X)$
in IIB and $K^{1}(X)$ in IIA string theories, where $X$ is the background
manifold.  On the other hand, world-volumes of Dp-branes correspond
to the cycles of K-homology groups, $K_{1}(X)$, $K_{0}(X)$, which
are dual to the $K$ theory groups. Let us see how K-cycles correspond
to the configurations of D-branes. 

A K-cycle on $X$ is a triple $(M,E,\phi)$ where $M$ is a compact
${\rm {Spin}^{c}}$ manifold without boundary, $E$ is a complex vector
bundle on $M$ and $\phi:M\to X$ is a continuous map. The topological
K-homology $K_{\star}(X)$ is the set of equivalence classes of the
triples $(M,E,\phi)$ respecting the following conditions:

\begin{itemize}

\item[(i)] $(M_{1},E_{1},\phi_{1})\sim(M_{2},E_{2},\phi_{2})$ when
there exists a triple (bordism of the triples) $(M,E,\phi)$ such
that $(\partial M,E_{|\partial M},\phi_{|\partial M})$ is isomorphic
to the disjoint union $(M_{1},E_{1},\phi_{1})\cup(-M_{2},E_{2},\phi_{2})$
where $-M_{2}$ is the reversed ${\rm {Spin}^{c}}$ structure of $M_{2}$
and $M$ is a compact ${\rm {Spin}^{c}}$ manifold with boundary. 

\item[(ii)] $(M,E_{1}\oplus E_{2},\phi)\sim(M,E_{1},\phi)\cup(M,E_{2},\phi)$,

\item[(iii)] Vector bundle modification $(M,E,\phi)\sim(\widehat{M},\widehat{H}\otimes\rho^{\star}(E),\phi\circ\rho)$.
$\widehat{M}$ is even dimensional sphere bundle on $M$, $\rho:\widehat{M}\to M$
projection, $\widehat{H}$ is a vector bundle on $\widehat{M}$ which
gives the generator of $K(S_{q}^{2p})=\mathbb{Z}$ on every $S_{q}^{2p}$
over each $q\in M$ \cite{Szabo2002a}. 

\end{itemize}

The topological K-homology defined above has an abelian group structure
where the sum is the disjoint union of cycles. The triples $(M,E,\phi)$
with $M$ of even dimension determines $K_{0}(X)$. Similarly, $K_{1}(X)$
corresponds to odd dimensions of $M$. Thus $K_{\star}(X)$ decomposes
into a direct sum of abelian groups:

\[
K_{\star}(X)=K_{0}(X)\oplus K_{1}(X)\,.\]
K-homology is dual to K-theory and the decomposition of $K_{*}(X)$
is a direct consequence of Bott periodicity (see \cite{Ati:67}).

Now one can interpret the cycles $(M,E,\phi)$ as D-branes \cite{HarveyMoore2000}:
$M$ is the world-volume of the brane, $E$ the Chan-Paton bundle
on it and $\phi$ gives the embedding of the brane into the (background)
spacetime $X$. Moreover, $M$ has to wrap the ${\rm Spin}^{c}$ manifold
\cite{FreedWitten1999} and $K_{0}(X)$ classifies stable D-branes
configurations in IIB, and $K_{1}(X)$ in IIA, string theories. The
equivalences of K-cycles as formulated in the conditions (i)-(iii)
correspond to natural relations for D-branes \cite{AsakawaSugimotoTerasima2002,Szabo2008b}. 

The topological K-homology theory above can be obtained analytically
(analytic K-homology theory). This theory is a special, commutative,
case of the following construction on general $C^{\star}$ algebras
\cite{AsakawaSugimotoTerasima2002}: A Fredholm module over a $C^{\star}$
algebra ${\cal A}$ is a triple $({\cal H},\phi,F)$ such that 
\begin{enumerate}
\item ${\cal H}$ is a separable Hilbert space,
\item $\phi$ is a $^{\star}$ homomorphism between $C^{\star}$ algebras
${\cal A}$ and ${\rm {\bf B}}({\cal H})$ of bounded linear operators
on ${\cal H}$,
\item $F$ is self-adjoint operator in ${\rm {\bf B}}({\cal H})$ satisfying
\end{enumerate}
\[
F^{2}-id\in{\rm K}({\cal H})\,,\quad[F,\phi(a)]\in{\rm K}({\cal H})\:{\rm for}\:{\rm every}\: a\in{\cal A}\]
where ${\rm K}({\cal H})$ are compact operators on ${\cal H}$. Now
let us see how a Fredholm module $({\cal H},\phi,F)$ describes certain
configuration of IIA K-matrix string theory related to D branes. To
this end we consider the operators of the K-matrix theory $\Phi^{0},...,\Phi^{9}$
(infinite matrices) acting on the Hilbert space ${\cal H}$ as generating
the $C^{\star}$ algebra ${\cal A}_{M}$ \cite{AsakawaSugimotoTerasima2002}.
In the case of commuting $\Phi^{\mu}$, hence commutative ${\cal A}_{M}$,
we have the following correspondence (explaining the index $M$ in
${\cal A}_{M}$): 
\begin{itemize}
\item Every commutative $C^{\star}$ algebra is isomorphic to the algebra
of continuous complex functions vanishing at infinity $C(M)$ on some
locally compact Hausdorff space $M$ (Gelfand-Naimark theorem and
Gelfand representation). A point $x\in M$ is determined by a character
of ${\cal A}_{M}$ which is a $^{\star}$ homomorphism $\phi_{x}:{\cal A}_{M}\to\mathbb{C}$.
\item $M$ serves as a common spectrum for $\Phi^{0},...,\Phi^{9}$ and
the choice of a point in $M$ represented as the eigenvalue of $\Phi^{\mu}$
fixes the position of the non BPS instanton along $x^{\mu}$.
\item In this way $M$ is covered by the positions of infinite many non
BPS instantons and serves as the world-volume of some higher dimensional
D brane \cite{AsakawaSugimotoTerasima2002}.
\end{itemize}
Now let us explain the role of the tachyon $T$. $T$ is a self-adjoint
unbounded operator acting on the Chan-Paton Hilbert space ${\cal H}$.
${\cal A}_{M}$ is a $C^{\star}$ unital algebra generated by $\Phi^{0},...,\Phi^{9}$
which can be now a noncommutative algebra. The corresponding geometry
of the world-volume $M$ would be a noncommutative geometry (in the
sense of Connes) and given by some spectral triple. The spectral triple
is in fact $({\cal H},{\cal A},T)$ which means that the following
conditions are fulfilled \cite{AsakawaSugimotoTerasima2002}:

\[
T-\lambda\in{\rm {\bf K}}({\cal H})\:\mbox{for every}\:\lambda\in\mathbb{C}\setminus\mathbb{R},\;[a,T]\in{\bf B}({\cal H})\:\mbox{for every}\: a\in{\cal A}_{M}\]
These conditions are fulfilled in our case of K-matrix string theory
for a tachyon field $T$, Chan-Paton Hilbert space ${\cal H}$ and
$C^{\star}$ algebra ${\cal A}_{M}$ generated by $\Phi^{\mu}$ .
Thus the natural extension of the spacetime manifold as well D-brane
world-volumes toward a noncommutative algebra and noncommutative world-volumes
of branes (represented by spectral triples) can be described by (see
e.g. \cite{AsakawaSugimotoTerasima2002}):
\begin{enumerate}
\item Fix the (spacetime) $C^{\star}$ algebra ${\cal A}$;
\item A $^{\star}$ homomorphism $\phi:{\cal A}\to{\bf B}({\cal H})$ generates
the embedding of the D-brane world-volume $M$ and its noncommutative
algebra ${\cal A}_{M}$ as ${\cal A}_{M}:=\phi({\cal A})$;
\item D-branes embedded in a spacetime ${\cal A}$ are represented by the
spectral triple $({\cal H},{\cal A}_{M},T)$;
\item Equivalently, a D-brane in $A$ is given by an unbounded Fredholm
module $({\cal H},\phi,T)$.
\end{enumerate}
Thus the classification of stable D-branes in ${\cal A}$ is given
by the classification of Fredholm modules $({\cal H},\phi,T)$ using
analytical K-homology. In the particular case of commutative $C^{\star}$
algebras based on the isomorphism of the topological and analytical
K-homology groups, we have the classification of stable D-branes in
terms of K-cycles, as was already discussed. In terms of K-matrix
string theory we can say that stable configurations of D-instantons
determine the stable higher dimensional D-branes which are K-homologically
classified as above \cite{AsakawaSugimotoTerasima2002}. 

Now let us turn to a more general situation than K-string theory of
D-instantons, i.e. backgrounds given by non-BPS Dp-branes or non-BPS
Dp-${\rm \overline{{\rm Dp}}}$-branes in type II string theory. Then
the stable configurations of Dq-branes are classified by generalized
K-theory namely Kasparov KK-theory \cite{AsakawaSugimotoTerasima2002}.
As in the case of D-branes in a $C^{\star}$ algebra ${\cal A}$ corresponding
to Fredholm modules, one defines an odd Kasparov module $({\cal H}_{{\cal B}},\phi,T)$,
where ${\cal H}_{{\cal B}}$ is an countable Hilbert module over the
$C^{\star}$algebra ${\cal B}$, by
\begin{itemize}
\item a $\star$-homomorphism from ${\cal A}$ to the $C^{\star}$ algebra
of bounded linear operators on ${\cal H}_{{\cal B}}$, $\phi:{\cal A}\to{\rm {\bf B}}({\cal H}_{{\cal B}})$;
\item a self-adjoint operator $T$ from ${\rm {\bf B}}({\cal H}_{{\cal B}})$
satisfying: 
\end{itemize}
\[
T^{2}-1\in{\rm {\bf K}}({\cal H}_{{\cal B}})\:\mbox{and}\:[T\,,\phi(a)]\in{\rm {\bf K}}({\cal H}_{{\cal B}})\:\mbox{for every}\, a\in{\cal A}\,,\]
where ${\rm {\bf K}}({\cal H}_{{\cal B}})$ is ${\cal B}\otimes{\bf {\rm K}}$.
$({\cal H}_{{\cal B}},\phi,T)$ is in fact a family of Fredholm modules
on the algebra ${\cal B}$. When ${\cal B}$ is $\mathbb{C}$ we have
an ordinary Fredholm module as before. The homotopy equivalence classes
of odd Kasparov modules $({\cal H}_{{\cal B}},\phi,T)$ determine
elements of $KK^{1}({\cal A},{\cal B})$. Also one defines even Kasparov
classes $KK^{0}({\cal A},{\cal B})=KK({\cal A},{\cal B})$ as homotopy
equivalence classes of the triples $({\cal H}_{{\cal B}}^{(0)}\oplus H_{{\cal B}}^{(1)},\phi^{(0)}\oplus\phi^{(1)},\left(\begin{array}{cc}
0 & T^{\star}\\
T & 0\end{array}\right))$. A natural $\mathbb{Z}_{2}$ grading appears due to the involution
${\cal H}_{{\cal B}}^{(0)}\oplus H_{{\cal B}}^{(1)}\to{\cal H}_{{\cal B}}^{(0)}\oplus-H_{{\cal B}}^{(1)}$. 

Now one obtains the classification pattern for branes in spaces. Let
us introduce non-BPS unstable Dp-branes wrapping the $p+1$-dimensional
world-volume $B$. Then stable Dq-branes configurations embedded in
a space $A$ transverse to $B$ corresponding to (are classified by)
the classes of $KK^{1}(A,B)$ (we identify the commutative algebras
$C(A)$, $C(B)$ with $A$, $B$ correspondingly). Similarly, given
non-BPS unstable Dp-${\rm \overline{Dp}}$-branes system, then stable
Dq-branes embedded in $A$ transverse to $B$ ($p+1$-dimensional
world-volumes) are classified by elements of $KK^{0}(A,B)$. The case
of even $KK^{0}(A,B)$ contains the $\mathbb{Z}_{2}$ grading as corresponding
to the Chan-Paton indices of Dp and ${\rm \overline{Dp}}$-branes.

\subsection{D-branes on separable $C^{\star}$ algebras and KK-theory \label{sub:Branes-on-separable}}

Thus the classification of D-branes in a spacetime manifold is given
by KK-theory as sketched in the previous subsection. This can be extended
over noncommutative spacetimes and noncommutative D-branes both represented
by separable $C^{\star}$ algebras as already can be seen from the
appearance of tools of KK-theory. First let us reformulate the {}``classic''
case of spaces in a way allowing this extension \cite{Szabo2008c}. 

In case of type II superstring theory, let $X$ be a compact part
of a spacetime manifold, i.e. $X$ is a compact ${\rm spin}^{c}$
manifold again with no background $H$-flux. As we saw, a flat D-brane
in $X$ is a Baum-Douglas K-cycle $(W,E,f)$. Here $f:W\hookrightarrow X$
is the embedding of the closed ${\rm spin}^{c}$ submanifold $W$
of $X$ and $E\to W$ is a complex vector bundle with connection (Chan-Paton
gauge bundle). It follows from the Baum-Douglas construction that
$E$ determines the stable class in the K-theory group $K^{0}(W)$
and all K-cycles form an additive category under disjoint union. Now,
the set of all K-cycles classes up to a kind of gauge equivalence
as in Baum-Douglas construction, gives the K-homology of $X$. This
K-homology is also the set of stable homotopy classes of Fredholm
modules taken over the commutative $C^{\star}$ algebra $C(X)$ of
continuous functions on $X$. This defines the correspondence (isomorphism)
between a K-cycle $(W,E,f)$ and the unbounded Fredholm module $({\cal H},\rho,D_{E}^{\mbox{W}})$.
Here ${\cal H}$ is the separable Hilbert space of square integrable
spinors on $W$ taking values in the bundle $E$, i.e. $L^{2}(W,S\otimes E)$,
$\rho:C(X)\to{\rm {\bf B}}({\cal H})$ is the representation of the
$C^{\star}$ algebra $C(X)$ in ${\cal H}$ such that $C(X)\ni g\to a_{g\circ f}\in{\rm {\bf B}}({\cal H})$
where $a_{g\circ f}$ is the operator of point-wise multiplication
of functions in $L^{2}(W,S\otimes E)$ by the function on $W$, $g\circ f$,
and $f:W\hookrightarrow X$. $D_{E}^{W}$ is the Dirac operator twisted
by $E$ corresponding to the ${\rm spin}^{c}$ structure on $W$.
Given the K-theory class of the Chan-Paton bundle $E$, i.e. $[E]\in K^{0}(W)$,
then the dual K-homology class of a D-brane, $[W,E,f]$ uniquely determines
$[E]$. In that way D-branes determine K-homology classes on $X$
which are dual to K-theory classes from $K^{r}(X)$ where $r$ is
the transversal dimension for the brane world-volume $W$. This K-theory
class is derived from the image of $[E]\in K^{0}(W)$ by the Gysin
K-theoretic map $f_{!}$. As we discussed already, the odd and even
classes of K-homology $K_{\star}(X)$ correspond to the parity of
the dimension of $W$. The K-cycle $(W,E,f)$ corresponds to a Dp-brane
and its gauge equivalence is given by Baum-Douglas construction using
the conditions (i)-(iii) in section \ref{sub:D-branes-on-spaces:}.
Thus we have \cite{Szabo2008b}:

Fact 1: \emph{There is a one-to-one correspondence between flat D-branes
in $X$, modulo Baum-Douglas equivalence, and stable homotopy classes
of Fredholm modules over the algebra $C(X)$.}

In the presence of a non-zero $B$-field on $X$, which is a $U(1)$-gerbe
with connection represented by the characteristic class in $H^{3}(X,\mathbb{Z})$
\cite{Szabo2008b,AsselmeyerKrol2009}, one can define twisted D-brane
on $X$ as \cite{Szabo2008b}:

\begin{definition}

A twisted D-brane in a B-field $(X,H)$ is a triple $(W,E,\phi)$,
where $\phi:W\hookrightarrow X$ is a closed, embedded oriented submanifold
with $\phi^{\star}H={\rm W}_{3}(W)$, and $E$ is the Chan-Paton bundle
on $W$, i.e. $E\in K^{0}(W)$, and ${\rm W}_{3}(W)$ is the 3-rd
integer Stiefel-Whitney class of the normal bundle of $W$, ${\rm W}_{3}(W)\in H^{3}(W,\mathbb{Z})$. 

\end{definition}

By the cancellation of the Freed-Witten anomaly, the condition in
the definition is really necessary. Let $H\in H^{3}(X,\mathbb{Z})$
represents the NS-NS $H$-flux. Since ${\rm W}_{3}(W)$ is the obstruction
to the existence of a ${\rm spin}^{c}$ structure on $W$ \cite{HiHo:58},
in the case of ${\rm W}_{3}(W)=0$ one has flat D-branes in $X$.
Thus equivalence classes of twisted D-branes on $X$ are represented
by twisted topological K-homology $K_{\star}(X,H)$ which is dual
to the twisted K-theory $K^{\star}(X,H)$. 

Now, in the case of $S^{3}$ and integral classes $H\in H^{3}(S^{3},\mathbb{Z})$,
one has some exotic $\mathbb{R}^{4}$'s which is determined by the
class $H$, when $S^{3}$ is a part of the boundary of the Akbulut
cork \cite{AsselmeyerKrol2010}. This is the same class $H$ which
twists the K-theory leading to $K^{\star}(S^{3},H)$. We can also
represent the $U(1)$ gerbes with connection on $S^{3}$, by the bundles
${\cal E}_{H}$ of algebras over $S^{3}$, such that the sections
of the bundle ${\cal E}_{H}$ define the noncommutative, twisted algebra
\emph{$C_{0}(X,{\cal E}_{H})$. }The Dixmier-Douady class of ${\cal E}_{H}$,
$\delta_{H}({\cal E}_{H})$, is again $H\in H^{3}(S^{3},\mathbb{Z})$
\cite{AsselmeyerKrol2009a,AtiyahSegal2004,Szabo2002a}. 

The important relation is the following (\cite{Szabo2008b}, Proposition
1.15): 

Fact 2: \emph{There is a one-to-one correspondence between twisted
D-branes in $(X,H)$ and stable homotopy classes of Fredholm modules
over the algebra $C_{0}(X,{\cal E}_{H})$.}

Since the algebra \emph{$C_{0}(X,{\cal E}_{H})$} determines its stable
homotopy classes of the Fredholm modules on it, then in the case $X=S^{3}$
one has the correspondence:

A. \emph{Let the exotic smooth $\mathbb{R}^{4}$'s are determined
by the integral third classes $H\in H^{3}(S^{3},\mathbb{Z})$. Then,
these exotic smooth $\mathbb{R}^{4}$'s correspond one-to-one to the
sets of twisted D-branes in $(S^{3},H)$, provided $S^{3}$ is a part
of the boundary of the Akbulut cork.}

Thus, given the complete collection of twisted D-branes in $(S^{3},H)$,
which take values in $K_{\star}(S^{3},H)$, one can determine, in
principle, the corresponding exotic $\mathbb{R}^{4}$. This exotic
$\mathbb{R}_{H}^{4}$ corresponds to the class $[H]\in H^{3}(S^{3})$
and the class $H$ twists the K-homology as dual to the twisted K-theory
$K^{\star}(S^{3},H)$ \cite{AsselmeyerKrol2009a,AsselmeyerKrol2010,Szabo2002a}.
In the following we try to convince the reader that the correspondence
of D-branes to 4-exotics can be extended to more general cases with
a closer relation. 

Remembering that $S^{3}\subset\mathbb{R}^{4}$ is a part of the Akbulut
cork of the exotic structure, our previous observation can be restated
as:

B. \emph{The change of the exotic smoothness of $\mathbb{R}^{4}$,
$\mathbb{R}_{H_{1}}^{4}\to\mathbb{R}_{H_{2}}^{4}$, $H_{1}$, $H_{2}\in H^{3}(S^{3},\mathbb{Z})$,
$H_{1}\neq H_{2}$, corresponds to the change of the curved backgrounds
$(S^{3},H_{1})\to(S^{3},H_{2})$ hence the sets of stable D-branes.}

This motivates the formulation:

C. \emph{Some small exotic smoothness appearing on $\mathbb{R}^{4}$,
$\mathbb{R}_{H_{1}}^{4}$, can destabilize (or stabilize) D-branes
in $(S^{3},H_{2})$, where $S^{3}\subset\mathbb{R}^{4}$ lies at the
boundary of the Akbulut cork of $\mathbb{R}_{H_{1}}^{4}$. We say
that D-branes configurations in $(S^{3},H_{2})$ are }4-exotic-sensitive\emph{.}

Next we extend the formalism of D-branes in spaces to \emph{quantum
}D-branes in general $C^{\star}$ algebras including the correspondence
described above,.\emph{ }There were developed recently impressive
counterparts of a variety of topological, geometrical and analytical
results, like Poincar\'e duality, characteristic classes and the
Riemann-Roch theorem, in $C^{\star}$ algebras. Besides the generalized
formula for charges of quantum D-branes in a noncommutative separable
$C^{\star}$ algebras was worked out \cite{Szabo2008a,Szabo2008b}.
Thus one obtains a suitable framework for considering the quantum
regime of D-branes. Therefore we will try to find a relation to 4-exotics
also in this quantum regime of D-branes. 

Following \cite{AsakawaSugimotoTerasima2002,Szabo2008a,Szabo2008b,Szabo2008c}
one can choose an obvious substitute for the category of quantum D-branes:
the category of separable $C^{\star}$ algebras where the morphisms
are elements of some KK-theory group. This means that for a pair $({\cal A},{\cal B})$
of separable $C^{\star}$ algebras the morphisms $h:{\cal A}\to{\cal B}$
is lifted to the element of the group $KK({\cal A},{\cal B})$. Thus
we can consider a generalized D-brane in a separable $C^{\star}$
algebra ${\cal A}$ as corresponding to the lift $h!:{\cal A}\to{\cal B}$
where ${\cal B}$ represents a quantum D-brane.

More precisely following \cite{Szabo2008a}, let us consider a subcategory
${\cal C}$ of the category of $C^{\star}$ separable algebras and
their morphisms, which consists of strongly K-oriented morphisms.
Therefore there exists a contravariant functor $!:{\cal C}\to KK$
such that ${\cal C}\ni f:{\cal A}\to{\cal B}$ is mapped to $f!\in KK_{d}({\cal B},{\cal A})$,
here $KK$ is the category of separable $C^{\star}$ algebras with
KK classes as morphisms. Strongly K-oriented morphisms and the functor
$!$ are subjects to the following conditions:
\begin{enumerate}
\item Identity morphism $id_{{\cal A}}:{\cal A}\to{\cal A}$ is strongly
K-oriented (SKKO) for every separable $C^{\star}$ algebra ${\cal A}$
and $(id_{{\cal A}})!=1_{{\cal A}}$. Also, the 0-morphism $0_{{\cal A}}:{\cal A}\to{\cal A}$
is SKKO and $(0_{{\cal A}})!=0\in KK(0,{\cal A})$.
\item If $f:{\cal A}\to{\cal B}$ is SKKO then $f^{\circ}:{\cal A}^{\circ}\to{\cal B}^{\circ}$
is also SKKO, and $(f!)^{\circ}=(f^{\circ})!$ where ${\cal A}^{\circ}$
is the opposite $C^{\star}$ algebra to ${\cal A}$, i.e. the algebra
having the same underlying vector space but the reversed product. 
\item Any morphism $f:{\cal A}\to{\cal B}$ is SKKO, provided ${\cal A}$
and ${\cal B}$ are strong Poincar\'e dual (PD) algebras. Then $f!$
is determined as: \begin{equation}
f!=(-1)^{d_{{\cal A}}}\Delta_{{\cal A}}^{\vee}\otimes_{{\cal A}^{0}}\left[f^{0}\right]\otimes_{{\cal B}^{0}}\Delta_{{\cal B}}\label{eq:K-orientation}\end{equation}
here $[f]$ is the class of $f:{\cal A}\to{\cal B}$ in $KK({\cal A},{\cal B})$.
$\Delta_{{\cal A}}$ is the fundamental class in $KK_{d_{{\cal A}}}({\cal A}\otimes{\cal A}^{\circ},\mathbb{C})=K^{d_{{\cal A}}}({\cal A}\otimes{\cal A}^{\circ})$,
$\Delta_{{\cal A}}^{\vee}$ its dual class in $KK_{-d_{{\cal A}}}(\mathbb{C},{\cal A}\otimes{\cal A}^{\circ})=K_{-d_{{\cal A}}}(A\otimes{\cal A}^{\circ})$
which exists by strong PD \cite{Szabo2008a}. 
\end{enumerate}
K-orientability was introduced in its original form, by A. Connes
to define the analogue of the ${\rm spin}^{c}$ structure for noncommutative
$C^{\star}$ algebras (see also \cite{Connes1984} and the next section).
The formulation of K-orientability and strong PD $C^{\star}$ algebras
are crucial ingredients of noncommutative versions of Riemann-Roch
theorem, Poincar\'e-like dualities, Gysin K-theory map and allows
to formulate a very general formula for noncommutative D-brane charges
\cite{Szabo2008b,Szabo2008a,Szabo2008c}. Let us notice that if both
${\cal A}$ and ${\cal B}$ are PD algebras then any morphism $f:{\cal A}\to{\cal B}$
is K-oriented and the K-orientation for $f$ is given in (\ref{eq:K-orientation}). 

In the special case of the proper smooth embedding $f:W\to X$ of
codimension $d$ between the smooth compact manifolds $W,X$, we choose
the normal bundle $\tau$ over $W$ to be ${\rm spin}^{c}$, where
$\tau$ is given by $TX=\tau\oplus f_{*}(TW)$. When $X$ is also
${\rm spin}^{c}$ then the ${\rm spin}^{c}$ condition on $\tau$
for vanishing $H$-flux in type II string theory formulated on $X$
is the Freed-Witten anomaly cancellation condition \cite{Szabo2008a}.
In this case any D-brane in $X$, given by the triple $(W,E,f)$,
determines the KK-theory element $f!\in KK(C(W),C(X))$. The construction
of K-orientation $f:M\to X$, between smooth compact manifolds, can
be extended to smooth proper maps which are not necessary embeddings.
Thus the general condition for K-orientability gives the correct analogue
for stable D-branes in $C^{\star}$ algebras. 

\begin{definition}\label{enu:Def: q-Branes}

\emph{A generalized stable quantum D-brane} on a separable $C^{\star}$
algebra ${\cal A}$, represented by a separable $C^{\star}$ algebra
${\cal B}$, is given by the strongly K-oriented homomorphism of $C^{\star}$
algebras, $h_{{\cal B}}:{\cal A}\to{\cal B}$. The K-orientation means
that there is the lift $(h_{{\cal B}})!\in KK({\cal B},{\cal A})$
where $!$ fulfills the functoriality condition as in (\ref{eq:K-orientation}).

\end{definition}

This approach to quantum D-branes is a natural extension of the string
formalism over $C^{\star}$algebras replacing spaces and branes, which
is currently a conjectural framework. This framework exceeds both
the dynamical Seiberg-Witten limit of superstring theory (inducing
noncommutative brane world-volumes) and the geometrical understanding
of branes, and places itself rather in a deep quantum regime of the
theory \cite{Szabo2008c}. On the other hand in such a formal quantum
limit of string theory one can observe the relation with 4-dimensional
exotic open smooth structures, which relies on the natural relation
of exotic $R^{4}$ with $C^{\star}$algebras of the foliations.

\section{Exotic $\mathbb{R}^{4}$ and branes in $C^{\star}$ algebras\label{sec:Exotic--and-branes}}

\subsection{Exotic $\mathbb{R}^{4}$ and stable D-branes configurations on foliated
manifolds\label{sub:D-branes-on-foliated}}

Now we want to tackle the problem to describe stable states of D-branes
in a more general geometry than used for spaces, namely the geometry
of foliated manifolds. The interesting case for us is a codimension-1
foliation of the 3-sphere $S^{3}$ admitting a noncommutative geometry
as we will show now. In general, to every foliation $(V,F)$ one can
associate its noncommutative $C^{\star}$ algebra $C^{\star}(V,F)$,
on the other hand a foliation determines its holonomy groupoid $G$
and the topological classifying space $BG$. Both cases, topological
K-homology of $G$ and $C^{\star}$algebraic K-theory, are in fact
dual. Analogously to our previous discussion of branes as K-cycles
on $X$, let us start with K-homology of $G$ and define D-branes
as K-cycles in $G$:

A $K$ - cycle on a foliated geometry $X=(V,F)$ is a triple $(M,E,\phi)$
where $M$ is a compact manifold without boundary, $E$ is a complex
vector bundle on $M$ and $\phi:M\to BG$ is a smooth K-oriented map.
Due to the K-orientability in the presence of canonical $G$-bundle
$\tau$ on $BG$, the condition of ${\rm Spin}^{c}$ structure on
$M$ is lifted to the ${\rm Spin}^{c}$ structure on $TM\oplus\phi^{\star}\tau$
\cite{Connes1984}. 

The topological K-homology $K_{\star,\tau}(X)=K_{\star,\tau}(BG)$
of the foliation $(V,F)$ is the set of equivalence classes of the
above triples, where the equivalence respects the following conditions:

\begin{itemize}

\item[(i)] $(M_{1},E_{1},\phi_{1})\sim(M_{2},E_{2},\phi_{2})$ when
there is a triple (bordism of the triples) $(M,E,\phi)$ such that
$(\partial M,E_{|\partial M},\phi_{|\partial M})$ is isomorphic to
the disjoint union $(M_{1},E_{1},\phi_{1})\cup(-M_{2},E_{2},\phi_{2})$
where $-M_{2}$ is the reversed ${\rm {Spin}^{c}}$ structure of $TM_{2}\oplus\phi_{2}^{\star}\tau$
and $M$ is a compact manifold with boundary. 

\item[(ii)] $(M,E_{1}\oplus E_{2},\phi)\sim(M,E_{1},\phi)\cup(M,E_{2},\phi)$,

\item[(iii)] Vector bundle modification $(M,E,\phi)\sim(\widehat{M},\widehat{H}\otimes\rho^{\star}(E),\phi\circ\rho)$
similarly as in the case of manifolds. 

\end{itemize}

As in the case of spaces (manifolds) and the corresponding K-homology
groups representing stable D-branes of type II superstring theory
(see section \ref{sub:D-branes-on-spaces:}), we generalize stable
D-branes as being represented by the above triples in case of the
geometry of foliated manifolds . 

\begin{theorem}

The class of generalized stable D-branes on the $C^{\star}$ algebra
$C^{\star}(S^{3},F_{1})$ (of the codimension 1 foliation of $S^{3}$)
corresponding to the K-homology classes $K_{\star,\tau}(S^{3}/F)$
determines an invariant of exotic smooth $\mathbb{R}^{4}$ . 

\end{theorem}

The result follows from the fact that $K_{\star,\tau}(S^{3}/F)$ is
isomorphic to $K_{\star,\tau}(BG)$ \cite{Connes1984} and this determines
a class of stable D-branes in $(S^{3},F)$. The foliations $(S^{3},F)$
correspond to different smoothings on $\mathbb{R}^{4}$ \cite{AsselmeyerKrol2009}.
$\square$

\subsection{The net of exotic $\mathbb{R}^{4}$'s and quantum D-branes in $C^{\star}(S^{3},F)$\label{sub:Net-of-exotic}}

The extension of string theory and D-branes to general noncommutative
separable $C^{\star}$ algebras can be considered as an approach to
quantum D-branes where D-branes are also represented by noncommutative
separable $C^{\star}$ algebras. A category of D-branes in a quantum
regime, is the category of separable $C^{\star}$ algebras where the
morphisms are elements of KK-theory groups. For a pair $({\cal A},{\cal B})$
of separable $C^{\star}$ algebras the morphisms $h:{\cal A}\to{\cal B}$
belong to $KK({\cal A},{\cal B})$. Abstract quantum D-branes in a
separable $C^{\star}$ algebra ${\cal A}$ correspond to $\phi:{\cal A}\to{\cal B}$
where ${\cal B}$ is the algebra representing a quantum D-brane and
$\phi$ is a strongly K-oriented map. A general formula for RR charges
in the noncommutative setting was obtained for these branes in \cite{Szabo2008a,Szabo2008b}.

D-branes, as considered in the previous subsection, correspond to
the lifted KK-theory classes. That is, if the D-brane corresponds
to the triple $(M,E,f)$ and $f:M\hookrightarrow G=V/F$ is a K-oriented
map then $f!\in KK(M,V/F)$ represents the D-brane (see \cite{Connes1984})
. More generally (still following \cite{Connes1984}), given a K-oriented
map $f:X\to Y$ , one can define (under certain conditions) a push
forward map $f!$ in K-theory. The very important property of the
analytical group $K(V/F)$ of the foliation $(V,F)$ is its ,,wrong
way'' (Gysin) functoriality, i.e. one associates to each K-oriented
map $f:V_{1}/F_{1}\to V_{2}/F_{2}$ of leaf spaces an element $f!$
of the Kasparov group $KK(C^{\star}(V_{1};F_{1});C^{\star}(V_{2};F_{2}))$. 

Now given a small exotic $\mathbb{R}^{4}$, say $e_{1}$, embedded
in some small exotic $\mathbb{R}^{4}$, $e$, both are represented
by the $C^{\star}$ algebras of the codimension-1 foliations of $S^{3}$,
$C^{\star}(V_{1};F_{1})$ and $C^{\star}(V;F)$ respectively. The
embedding $i:e_{1}\hookrightarrow e$ determines the corresponding
K-oriented map of the leaf spaces $f_{i}:S^{3}/F_{1}\to S^{3}/F$
and the KK-theory lift $f_{i}!\in KK(C^{\star}(V_{1};F_{1});C^{\star}(V;F))$.
According to definition \ref{enu:Def: q-Branes} from section \ref{sub:Branes-on-separable},
we obtain

\begin{theorem}

\label{theo:quantum-exotic-R4}

Let $e$ be an exotic $\mathbb{R}^{4}$ corresponding to the codimension-1
foliation of $S^{3}$ which gives rise to the $C^{\star}$algebra
${\cal A}_{e}$. The exotic smooth $\mathbb{R}^{4}$ embedded in $e$
determines a generalized quantum D-brane in ${\cal A}_{e}$. 

\end{theorem}

Given exotic $\mathbb{R}^{4}$'s, $\{e_{a},\, a\in I\}$, all embedded
in $e$, one has the family of $C^{\star}$ algebras, $\{{\cal A}_{a},\, a\in I\}$,
of the codimension-1 foliations of $S_{a}^{3},\: a\in I$. Now the
embeddings $e_{a}\to e$ determine the corresponding K-oriented maps
of the leaf spaces as before, and the $\star$-homomorphisms of algebras
$\phi_{a}:{\cal A}_{e}\to{\cal A}_{a}$. The corresponding classes
in KK-theory $KK({\cal A}_{e},{\cal A}_{a})$ represent the quantum
D-branes in ${\cal A}_{e}$. $\square$

However, the correspondence in the theorem is many-to-one and an exotic
smooth $\mathbb{R}^{4}$ embedded in $e$ can be represented (non-uniquely)
by stable D-brane in ${\cal A}_{e}$, and not all abstract D-branes
in the algebra ${\cal A}_{e}$ are represented by some exotic $e'\subset e$.
Still one can consider D-branes represented by exotic $e_{a}$ in
$e$ as carrying 4-dimensional, hence potentially physical, information.
This is a kind of special ,,superselection'' rule in superstring theory
and will be discussed separately.

\section{Discussion and conclusions}

In this paper we give further evidences that string theory is indeed
related to 4-dimensional nonstandard smoothness of open manifolds
like $\mathbb{R}^{4}$. Our concern here was the quantum limit of
D-branes. We show that, on the formal level, there are strong correlations
between formalism of quantum D-branes in a quantum spacetime, both
represented by separable $C^{\star}$-algebras, and exotic smooth
$\mathbb{R}^{4}$'s. These $\mathbb{R}^{4}$'s are also represented
by $C^{\star}$-algebras and embedded in some exotic $\mathbb{R}^{4}$.
These $C^{*}-$algebras are the convolution algebras of the codimension
1 foliations of the 3-sphere when $S^{3}$ is taken as a part of the
boundary of the Akbulut cork for the small exotic $\mathbb{R}^{4}$.
Thus we model quantum D-branes in a quantum spacetime by exotic $\mathbb{R}^{4}$'s
embedded in an exotic $\mathbb{R}^{4}$. When the target ,,spacetime''
$\mathbb{R}^{4}$ is taken to be the standard one, which is always
possible since exotic $\mathbb{R}^{4}$'s are \emph{small}, one recovers
the correlation with ,,classic'' configurations of D and NS branes
in certain string backgrounds, as was described in our previous paper
\cite{AsselmeyerKrol2011}. Thus the way to abstract algebraic setting
of $C^{\star}$-algebras and quantum D-branes generalizes the correspondence
of branes (represented by submanifolds or K-homology classes) with
exotic $\mathbb{R}^{4}$ seen as smooth submanifolds of the standard
$\mathbb{R}^{4}$. These two-facets of exotic $\mathbb{R}^{4}$, namely
the $C^{\star}$ algebraic and smooth (sub)manifold, are crucial for
exhibiting the full range of a correspondence to string theory. When
smoothness on $\mathbb{R}^{4}$ is standard we lose the string information
as encoded in 4-structures. Thus, one could in some important cases
translate stringy situations into 4-smooth setting and conversely
and this is not a duplicate of existing approaches in string theory.
Thus we gain the additional and independent channel leading to 4-dimensions
from string theory. The crucial is whether this 4-dimensional data
carry information on physics in 4-dimensions. This important point
was considered already in a series of research papers \cite{BraRan:93,Bra:94a,Bra:94b,Ass:96,AssBra:2002,Ass2010,Sla:96,Sladkowski2001}
as well in a textbook \cite{Asselmeyer2007}. In \cite{AsselmeyerKrol2009}
we showed that exotic smoothness of an open 4-region in spacetime
have the same effect as the existence of magnetic monopoles, i.e.
exotic smoothness induces the quantization condition for the electric
charge. Moreover, one can consider exotic $\mathbb{R}^{4}$'s as quantum
object, i.e. the spacetime induces the quantization processes \cite{AsselmeyerKrol2010}. 

However, the full-fledged presentation of the relation of exotic $\mathbb{R}^{4}$,
and other open smooth 4-manifolds, with string theory is out of reach
for the authors at present. We think that new analytical and topological
tools are needed. In the forthcoming paper we will present an effort
into this direction and try to understand quantum branes as a kind
of wild embeddings based on the smoothness of 4-manifolds. Thus the
point of the question ,,Is it possible that string theory deals with
4-dimensional structures directly neither by implementing compactifications
nor by phenomenological models-building, and these structures would
have a physical meaning?'' should be further explored and studied.
As we emphasized in the previous paper \cite{AsselmeyerKrol2011}
this effort should help with understanding both 4-dimensional physics
as appearing from string theory and exotic open 4-manifolds in mathematics
\cite{AssKrol2010ICM}.

\section*{Acknowledgment}

T.A. wants to thank C.H. Brans and H. Ros\'e for numerous discussions
over the years about the relation of exotic smoothness to physics.
J.K. benefited much from the explanations given to him by Robert Gompf
regarding 4-smoothness several years ago, and discussions with Jan
S{\l}adkowski.

%\bibliographystyle{plain}
%\addcontentsline{toc}{section}{\refname}\bibliography{knots,diffbib,foliation-gerbes,E:/torsten/MyPapers/bib/knots,E:/torsten/MyPapers/bib/diffbib,E:/torsten/MyPapers/bib/foliation-gerbes,foliation-gerbes-1}

\end{document}